\documentclass[11pt]{article}
\usepackage{graphicx}
\usepackage{amsmath}
\usepackage{amsfonts}
\usepackage{amssymb}
\usepackage{epsfig}
\usepackage{times}
\usepackage{calc}
\usepackage{version}
\usepackage[french,english]{babel}
\def\la{\langle}
\def\ra{\rangle}
\begin{document}


\begin{titlepage}


\null

\vskip 4cm

{\bf\large\baselineskip 20pt
\begin{center}
Particle Multiplicity in Jets and Sub-jets with Jet Axis from Color Current
\end{center}
}
\vskip 1cm

\begin{center}
Wolfgang Ochs\footnote{E-mail: wwo@mppmu.mpg.de}\\
\smallskip
Max-Planck Institut f\"ur Physik,
Werner-Heinsenberg-Institut\\
F\"oringer Ring 6, D-80805 M\"unchen, Germany\\

\medskip
Redamy P\'erez Ramos\footnote{E-mail: redamy@mail.desy.de}\\
\smallskip
II. Institut f\"ur Theoretische Physik, Universit\"at Hamburg\\
Luruper Chaussee 149, D-22761 Hamburg, Germany
\end{center}

\baselineskip=15pt

\vskip 1cm

{\bf Abstract}: We study the particle multiplicity in a jet or sub-jet
as derived from an
energy-multiplicity 2-particle correlation. This definition 
avoids the notion of a
globally fixed jet axis and allows for the study of smaller jet cone openings
in a more stable way. The results are
sensitive to the mean color current $\langle C \rangle_{A_0} $ in the jet from
primary parton $A_0$ which takes into account intermediate 
partonic processes in the sub-jet production where
$C_F< \langle C \rangle_{A_0} < N_c$ at high energies. We 
generalize previous calculations in Leading Logarithmic Approximation (LLA).
The size of the effects related to this jet axis definition are computed for
multiplicities in sub-jets with different opening 
angles and energies by including 
contributions from the Modified LLA (MLLA)
and Next-to-MLLA to the leading order QCD results.
   
\vskip .5 cm

{\em Keywords: perturbative Quantum Chromodynamics, jets, multiplicity}

\vfill


\end{titlepage}




\section{Introduction}

\label{section:intro}

The collimation of hadrons in jets is a basic phenomenon of high energy
collisions and its quantitative understanding is an important task for QCD.
Simple differential characteristics of a jet are their energy and
multiplicity angular profiles. 
The collimation of energy and multiplicity in the jet
follows from the dominance of gluon bremsstrahlung processes in the parton
cascade evolution. Whereas the former is more sensitive to the hard 
processes inside a jet the second one is sensitive to the soft parton emissions
from the primary parton. 

The characteristics of soft particle production, such as particle
multiplicities, inclusive distributions and correlation functions, are
derived in QCD in the Modified Leading Logarithmic Approximation (MLLA) (for
review, see \cite{Basics}) which takes into account the leading double
logarithmic terms and the single logarithmic corrections. These azimuthally
averaged quantities can be obtained from an evolution equation for the
generating functional of the parton cascade. This equation provides also
Next-to-MLLA corrections taking into account energy conservation of parton
splittings with increased accuracy. The corresponding hadronic observables
can be obtained using the concept of Local Parton Hadron Duality (LPHD)
\cite{LPHD} which has turned out a successful description of many hadronic
phenomena (see, e.g. \cite{KhozeOchs}).
 
A characteristic prediction of QCD is the increased 
mean particle multiplicity in a gluon
jet over the quark jet, asymptotically by the ratio of color factors
 $r=N_G/N_Q \to N_c/C_F=9/4$ modified by corrections in powers of
$\sqrt{\alpha_s}$ \cite{MultTheory,MALAZA,DREMIN}. These results have been compared with LEP data on 2- and
3-jet events, interpreted as $q\bar q$ and $q\bar q g$ primary production
\cite{OPALrgq,DELPHIrgq,DreminGary}.
In applications with primary hadron beams, such as at HERA, RHIC, 
TEVATRON or LHC
it is necessary to select jets by an angular cut to remove the effects from
the hadron remnants. Such an analysis of multiplicities has been performed
at the TEVATRON \cite{TevMult} with different jet opening angles $\Theta$ and
energies $E$. The multiplicity as well as differential multiplicity
distributions in MLLA 
depend on the maximum transverse momentum in the jet (``virtuality''), 
i.e. on the combination
\begin{equation}\label{eq:scale}
Q\approx E\Theta 
\end{equation}
in the small angle
approximation. This scaling behavior has been verified \cite{TevMult} for the
multiplicities $N_G,\ N_Q$ in the angular range $\Theta=0.28\ldots0.47$ 
and also for the peak position of the 
inclusive rapidity spectra \cite{TevRap}.
Results from LEP and TEVATRON support in general 
the theoretical treatment based on the MLLA evolution 
equation and the agreement with data improves with the
inclusion of higher order terms in $\sqrt{\alpha_s}$.

For small opening angles $\Theta$ the 
definition of the jet axis becomes problematic because of the fluctuations
at low particle multiplicity.
The multiplicity at smaller angles $\Theta$  
can be defined in a more stable way without reference to a jet axis
by the double
inclusive correlation density $N_2$ between two particles.
The particles (p2) within the cone
$\Theta$ around particle (p1) are counted 
with angle $\Theta_{12}$ (including particle (p1) itself), 
then the superposition of the
particles (p1) in the jet weighted by their energies $E_1$ is performed.
For a jet of full energy $E$, in a form differential 
in the energy fraction $x_2$, 
 the multiplicity in the cone $\Theta$ is given as (see Ref. \cite{Basics})
 \begin{eqnarray}
\frac{dN_2(x_2,E,\Theta)}{dx_2} &=& 
    \frac{1}{E}\int_0^\Theta d\Theta_{12} \int dE_1 E_1 \int dE_2
    \frac{d\sigma_2(E)}{\sigma_1 dE_1 dE_2 d\Theta_{12}} 
     \delta\left(x_2-\frac{E_2}{E}\right)
   \nonumber\\
 \sigma_1 &=& \int dE_1  \frac{E_1 d\sigma_1(E)}{dE_1}.
   \label{eq:mult2}
\end{eqnarray}
In this definition, the collection of particles (p1) 
serves to replace the jet axis and therefore, also small angles are well
defined, which is not the case for the usual global jet axis definition.    
 
The evaluation of Eq. (\ref{eq:mult2}) for the distribution in energy $x_2$ 
and, after integration, for the multiplicity of particles (p2) in the cone
has been obtained in LLA for the integral over the energy $E_1$ 
in Refs. \cite{DDT,Basics}.
The definition of jet axis is especially relevant for the inclusive
distribution of the transverse momentum 
$k_T$ of particles and this quantity has been derived 
with reference to the energy
flux in the MLLA
\cite{PerezMachet} and including 
an important class of Next-to-MLLA
(NMLLA) contributions as obtained from the evolution equation \cite{PAM}. 
The predictions on the $k_T$ distributions have been found in a rather good
agreement with the preliminary data ($k_T>1$ GeV) from the CDF 
collaboration \cite{CDF} using so far a conventional jet definition.
In this paper we study the effects from the definition
Eq. (\ref{eq:mult2}) and derive results for particles confined into small
cones. This allows the study of particle distributions with smaller $k_T$ in
the jet and lower effective energies (virtuality) $E\Theta$ than the nominal
jet energy $E$. 

For decreasing cone size $\Theta$ of a sub-jet in the definition of Eq.
(\ref{eq:mult2}) the simple scaling
behavior as in Eq. (\ref{eq:scale}) will be broken. The occurrence of
intermediate processes leads to a mixing of quark and gluon sources
and, asymptotically in the LLA, the hadron 
multiplicity $\hat N^h_{A_0}$ in the sub-jet with opening angle $\Theta$
in the primary jet from parton $A_0$ ($Q$ or $G$) and large opening angle 
$\Theta_0\sim 1$ is 
given by \cite{Basics} 
\begin{equation}
\hat N^h_{A_0}(\Theta,E,\Theta_0)=\frac{1}{N_c} \langle C\rangle_{A_0}\cdot N_G^h
(E\Theta),
\label{ColCurDef}
\end{equation}
proportional to the multiplicity $N_G^h$ of an isolated gluon jet of
virtuality $E\Theta$.
The average color current $\langle C\rangle_{A_0}$ 
inside the small cone is obtained in the same approximation \cite{Basics} as
\begin{equation}
\langle C\rangle_{A_0}=\langle u\rangle_{A_0}^G\cdot N_c 
    + \langle u\rangle_{A_0}^Q \cdot C_F,
\end{equation}
where $\langle u\rangle_{A_0}^A $ denotes the mean energy fraction of parton $A$
inside parton jet $A_0$ for given jet virtualities $E\Theta$ and $E\Theta_0$. 
This quantity is found in between the color factors
\begin{equation}
C_F< \langle C \rangle_{A_0} < N_c.
\label{inequality}
\end{equation}
The hadron multiplicity $N^h$, as suggested by LPHD, 
is taken proportional to the parton multiplicity 
$N$ at cut-off scale $Q_0$, i.e. $N^h(E\Theta)=K \cdot N(E\Theta,Q_0)$ 
with hadronization constant $K$. 

Using the alternative approach for jet definition from Eq.
(\ref{eq:mult2})
we derive in this paper results on the particle multiplicity as a simple
observable showing its dependence on the color
current and its variation with 
jet energy and opening angle $\Theta$
in the improved accuracy of MLLA and NMLLA. Some
phenomenological applications will be studied.


\section{Evolution equations for multiplicities in quark and gluon jets}
\label{section:multiplicity}
%
In MLLA the evolution of azimuthally averaged quantities with jet energy $E$
and jet opening $\Theta$ is given by an evolution equation for the
generating functional for the parton momenta in 
the jet \cite{Basics}. The evolution involves 
$\alpha_s$, the running coupling constant of QCD
\begin{equation}\label{eq:var}
\alpha_s\equiv\alpha_s(E\Theta)=\frac{2\pi}{4N_c\beta_0
\ln\left(\frac{E\Theta}{\Lambda}\right)}, \qquad
\beta_0=\frac1{4N_c}\left(\frac{11}3N_c-\frac43T_R\right)
\end{equation}
where $\Lambda\equiv\Lambda_{QCD}$ is the intrinsic scale of QCD and
$\beta_0$ is the first term in the perturbative expansion of the $\beta-$function
\footnote{Following common practice, we consider in this work on soft
particles
the 1-loop expression for the running coupling constant, therefore $\Lambda$
has to be considered a phenomenological parameter not related to the usual
$\Lambda_{\overline{MS}}$. 
\label{ftn:2loopalpha}},
$N_c$ is the number of colors, $T_R=n_f/2$, where $n_f$ is the number of
light quark flavors. In Double Logarithmic Approximation (DLA) $\alpha_s$ is
also linked with the anomalous dimension $\gamma_0$ of twist-2
operators by
\begin{equation}\label{eq:anodim}
\gamma_0^2\equiv\gamma_0^2(E\Theta) = 2N_c\frac{\alpha_s(E\Theta)}{\pi}=
\frac1{\beta_0(Y_{\Theta}+\lambda)};\quad
Y_{\Theta}=\ln\frac{E\Theta}{Q_0},\quad\lambda
=\ln\frac{Q_0}{\Lambda},
\end{equation}
where 
 $Q_0$ is the collinear cut-off parameter for $k_T=E\Theta>Q_0$. The results depend on energy and
angle only through the variable $Y_\Theta$, i.e. the maximum transverse
momentum in the jet.
We also set $Y'_\Theta=Y_\Theta+\lambda$ in the following.

For the evolution of the mean multiplicities in quark and gluon jets 
one obtains in MLLA
the coupled system of two evolution equations \cite{Basics} 
\begin{eqnarray}
\frac{d}{dY_\Theta}N_G^h(Y_\Theta)&=&\int_0^1 dx\,
\gamma_0^2(Y'_\Theta)\left[\Phi_G^G(x)\left(N_G^h(Y_\Theta+\ln x)+
N_G^h(Y_\Theta+\ln(1-x))-N_G^h(Y_\Theta)\right)\right.\nonumber \\
&+&\left.n_f\Phi_G^Q(x)\left(N_Q^h(Y_\Theta+\ln x)+
N_Q^h(Y_\Theta+\ln(1-x))-N_Q^h(Y_\Theta)\right)\right],\label{eq:NGh}\\
\frac{d}{dY_\Theta}N_Q^h(Y_\Theta)&=&\int_0^1 dx\,
\gamma_0^2(Y'_\Theta)\left[\Phi_Q^G(x)\left(N_G^h(Y_\Theta+\ln x)+
N_Q^h(Y_\Theta+\ln(1-x))-N_Q^h(Y_\Theta)\right)\right].\nonumber
\end{eqnarray}
Together with initial conditions for the multiplicities at threshold
(every jet contains only one parton, quark or gluon) these equations
determine the multiplicities at the higher values of $E\Theta$.
The DGLAP splitting functions denoted by  $\Phi_A^B$ are given by
$$
\Phi_G^G(x)=\frac{1}{x}-(1-x)[2-x(1-x)],\quad
\Phi_G^Q(x)=\frac{1}{4N_c}[x^2+(1-x)^2],\quad
\Phi_Q^G=\frac{C_F}{N_c}\left(\frac{1}{x}-1+\frac{x}{2}\right).
$$
The equations above for multiplicities
can be solved analytically \cite{DreminGary} in terms of
an expansion in $\gamma_0$ of the ratio of
multiplicities $r$ and the QCD
anomalous dimension $\gamma$
defined by
\begin{equation}
r=\frac{N_G^h}{N_Q^h},\qquad\gamma=\frac{d\ln N_G^h}{dY_\Theta}.
\label{eq:rgq}
\end{equation}

The Eqs. (\ref{eq:NGh}) 
are complete only up to 
MLLA order. The emerging higher order terms in the asymptotic expansion, 
although not complete, 
provide, however, an important contribution improving the constraint from
energy conservation. In particular, they build the correct threshold 
behavior of the 
parton cascade and therefore a better behavior at the present energies
is obtained. A full numerical solution of the evolution equation
\cite{LupiaOchs} corresponding to an all order resummation has in fact
provided a rather close reproduction of the multiplicity ratio $r$ above; the
same is also true for the result of the HERWIG MC \cite{HERWIG} at parton level and at
hadron level.
On the other hand, the truncated series in  $\gamma_0$ while asymptotically
convergent, diverges at low energies. 

We limit ourselves here to the NMLLA expression of $r$ that reads
\begin{equation}\label{eq:QGratio}
r=r_0(1-r_1\gamma_0-r_2\gamma_0^2)+{\cal O}(\gamma_0^3)
\end{equation} 
where the asymptotic value of $r$ is $r_0=N_c/C_F=9/4$. 
The MLLA term $r_1$ has been calculated in \cite{MultTheory}.
The coefficients $r_k$ can be
obtained from the Taylor expansions
of $N_i^h(Y_\Theta+\ln x)$ and $N_i^h(Y_\Theta+\ln (1-x))$
for $Y_\Theta\gg\ln x$ and $Y_\Theta\gg\ln(1-x)$ respectively and from
equating the terms of the same order in $\gamma_0$
in both sides of Eq. (\ref{eq:NGh});
 the values for $N_f=3$ for NMLLA
read $r_1=0.185$ and $r_2=0.426$ \cite{DreminGary}. The NMLLA solution
for the mean multiplicity in a gluon jet is found as \cite{DreminGary}
\begin{equation}\label{eq:NGhNMLLA}
N_G^h(Y_\Theta)\simeq K'\left(Y'_\Theta\right)^{-a_1/\beta_0}
\exp{\left(\frac{2}{\sqrt{\beta_0}}\sqrt{Y'_\Theta}-
\frac{2a_2}{\beta_0^{3/2}\sqrt{Y'_\Theta}}\right)} 
\end{equation}
with $a_1=0.28$, $a_2=0.38$ at $n_f=3$; $K$ is the LPHD
normalization factor.
The pre-exponential term $(Y'_\Theta)^{-a_1/\beta_0}$
is the MLLA contribution to $N_G^h$, while
the one $\propto a_2$, the NMLLA one.
We give accordingly, the ${\cal O}(\gamma_0^2)$
expressions of the first and second logarithmic derivatives of $N_G^h$
that follow from Eq. (\ref{eq:NGhNMLLA})
\begin{equation}\label{eq:logderN}
\frac{d\ln N_G^h}{dY_\Theta}=\gamma_0(E\Theta)
-a_1\gamma_0^2(E\Theta)+{\cal O}(\gamma_0^3),
\end{equation}
\begin{equation}\label{eq:seclogderN}
\frac{1}{N_G^h}\frac{d^2 N_G^h}{{dY}^2_\Theta}
=\left(\frac{d\ln N_G^h}{d{Y}_\Theta}\right)^2+
\frac{d^2\ln N_G^h}{{dY}^2_\Theta}=\gamma_0^2(E\Theta)+{\cal O}(\gamma_0^3)
\end{equation}
which we will all use in the following. The NMLLA solution for $N_Q^h$
can be obtained by substituting Eq. (\ref{eq:NGhNMLLA}) into Eq. 
(\ref{eq:QGratio}).

\section{Energy-multiplicity correlations}
\label{subsec:collimation}

We consider 
the production of a jet in a high energy collision ($pp$, $p\bar p$, $ep$,  
$e^+e^-$ \ldots)
initiated from a parton $A_0$ of energy $E$ 
and with opening angle $\Theta_0$ which
separates it from other jets.
\begin{figure}[h]
\begin{center}
\epsfig{file=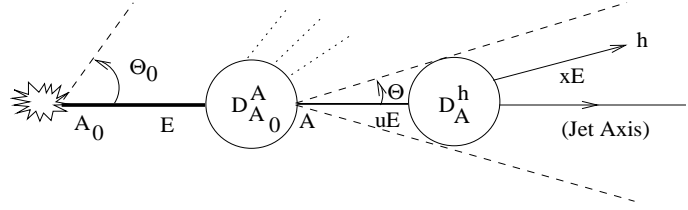, height=2.7truecm,width=9truecm}
\caption{\label{fig:distri} Inclusive production of hadron $h$ 
in a sub-jet of opening angle $\Theta$
inside a high energy jet of total opening angle $\Theta_0$.
The ``jet axis'' of parton $A$ corresponds to the
direction of the energy flux, i.e. the energy weighted direction 
of particles in the jet according to Eq. (\protect\ref{eq:mult2}).}  
\end{center}
\end{figure}
A sub-jet is defined by the opening angle $\Theta<\Theta_0$ 
with respect to parton $A$
with momentum fraction $u$. The distribution
$F_{A_0}^h(x,\Theta,E,\Theta_0)$ of the momentum fraction $x$ 
of a hadron $h$ in this sub-jet is
obtained by the integration of the 
double-inclusive correlation function in Eq. \ref{eq:mult2}, 
see Ref. \cite{DDT}. 
This process is depicted in Fig.~\ref{fig:distri}. 
Notations and kinematics are identical to the ones used in the work of
Ref. \cite{PerezMachet} which also contains further details.

The distribution $F_{A_0}^h$
is obtained as the convolution of two
fragmentation functions \cite{Basics}
\begin{equation}
 F_{A_0}^{h}\left(x,\Theta,E,\Theta_0\right)
= \sum_{A=Q,G}\int_x^1 du\,
 D_{A_0}^A\left(u,E\Theta_0,uE\Theta\right)D_A^{h}\left(\frac{x}
 {u},uE\Theta,Q_0\right).
\label{eq:F}
\end{equation}
This equation expresses the
correlation between the energy flux and one particle ($h$) within the sub-jet
with angle $\Theta$ generated from intermediate parton $A$ and with energy
fraction
$u$. The function
$D_{A_0}^A$ in (\ref{eq:F}) describes
the probability to emit parton $A$ of virtuality $uE\Theta$
with energy fraction $u$ off the initial parton $A_0$
($Q$ or $G$),  taking into account the evolution
of the jet between $Q=E\Theta_0$ and $Q=E\Theta$. 
The function $D_{A}^h$ in Eq. (\ref{eq:F}) describes the
probability to produce the hadron $h$ off $A$ with energy fraction $x/u$ and
transverse momentum scale $k_\perp\approx uE\Theta\geq Q_0$.

Integrating (\ref{eq:F}) over the energy fraction $x$ yields the
corresponding sub-jet multiplicity $\hat N_{A_0}^h$ of hadrons  
inside the angular range $\Theta<\Theta_0$ of the jet $A_0$
\begin{equation}\label{eq:Nconv}
\hat N_{A_0}^h(\Theta;E,\Theta_0)\approx\sum_{A=Q,G}\int_{Q_0/E\Theta}^1 du\,u\,
D_{A_0}^A\left(u,E\Theta_0,uE\Theta\right)
N_A^{h}\left(uE\Theta,Q_0\right),
\end{equation}
where $N_A^h$ is the number of hadrons (partons at scale $Q_0$)
produced inside the sub-jet $A$
of total virtuality $uE\Theta\geq Q_0$. 

We also note two limits of the sub-jet multiplicity in Eq.
(\ref{eq:Nconv}). At first, in the limit
of large angle $\Theta\to \Theta_0$ one finds, see also \cite{Basics},
\begin{equation}
\Theta\to \Theta_0: \quad D_{A_0}^A \to \delta(1-u)\delta_{A_0}^A, 
\quad \hat N_{A_0}^h(\Theta;E,\Theta_0)\to N_{A_0}^h(E\Theta_0;Q_0)
\label{subjetlim1} 
\end{equation}
and the sub-jet multiplicity coincides with the overall jet multiplicity.
At small opening angle $\Theta \to \frac{Q_0}{uE}$
\begin{equation}
\Theta\to 0: \quad D_{A}^h \to \delta(1-x/u)\delta_{A}^h,
\quad \hat N_{A_0}^h(\Theta;E,\Theta_0)\to 1.
\label{subjetlim2}
\end{equation}
the sub-jet multiplicity approaches the limit of one particle. This limit
cannot be reached in our approximation with a truncated expansion in
$\gamma_0$.

\subsection{Approximations with the leading parton}
\label{subsub:SA}

The convolution integral in Eq. (\ref{eq:Nconv})
is dominated by the region $u\approx1$ as $N_A^h(uE\Theta,Q_0)$
is exponentially rising with $u$ and the behavior
 $D_{A_0}^A(u)\sim u^{-1} \rho(\ln u)$ 
leads to a slowly varying function $uD_{A_0}^A(u)$
(for further discussion, see \cite{PerezMachet}). 
Therefore, we use the DGLAP expressions for
$D_{A_0}^A\left(u,E\Theta_0,uE\Theta\right)$ in the following, then the
evolution between the two scales depends only on the variable
\begin{equation}\label{eq:xiu}
\xi(u)=\frac1{4N_c\beta_0}\ln\left(\frac{\ln\frac{E\Theta_0}{\Lambda}}
{\ln\frac{uE\Theta}{\Lambda}}\right)\equiv\frac1{4N_c\beta_0}\ln
\left(\frac{Y_{\Theta_0}+\lambda}{\ln u+Y_\Theta+\lambda}\right).
\end{equation}
and for $D_{A_0}^A\left(u,E\Theta_0,uE\Theta\right)$ 
we also write $D_{A_0}^A\left(u,\xi(u)\right)$ for short.


Within the leading parton approximation
we expand the multiplicity
$N_A^h(uE\Theta,Q_0)$ in (\ref{eq:Nconv}) at $u\sim 1$. 
As in the MLLA evolution equation (\ref{eq:NGh}) and thereafter 
we take a logarithmic dependence of multiplicity $N_A^h$ and for
$\ln u\ll Y_\Theta\equiv\ln(E\Theta/Q_0)$ and 
for $E\Theta\gg \Lambda$, this quantity
can be written as
\begin{equation}
N_A^h\left(\ln u+Y_\Theta\right)
\stackrel{u\sim1}{\approx} N_A^h\left(Y_\Theta\right)+
\ln u\frac{dN_A^h\left(Y_\Theta\right)}{dY_\Theta}
+\frac{1}{2}\ln^2 u\frac{d^2N_A^h
\left(Y_\Theta\right)}{d{Y}_\Theta^2} + {\cal O}(\alpha_s^{3/2}).
\end{equation}
up to the  NMLLA level of accuracy.
Therefore, in this  approximation the correlation integral
in (\ref{eq:Nconv}) can be replaced by a sequence of factorized terms
showing alternating sign
\begin{equation}\label{eq:decorr}
\hat N_{A_0}^h(Y_{\Theta_0},Y_\Theta)\approx\sum_{A}\left(\la u\ra_{A_0}^A
N_A^{h}+\la u\ln u\ra_{A_0}^A\frac{d N_A^h}{dY_\Theta}
+\frac{1}{2}\la u\ln^2 u\ra_{A_0}^A\frac{d^2N_A^h}{d{Y}^2_\Theta}\right)
+{\cal O}(\alpha_s^{3/2}).
\end{equation}
Here $N_A^h$ depends only on $Y_\Theta$ and the mean energy fraction 
$\la u\ra_{A_0}^A$ is given by
\begin{equation}\label{eq:sv}
\la u\ra _{A_0}^A=\tilde D_{A_0}^A(j=2,\xi(1))+{\cal O}(\alpha_s),
\end{equation}
where $\tilde D_{A_0}^A(j,\xi(1))$ is the representation of DGLAP
fragmentation functions in Mellin's space \cite{Basics}
\begin{equation}
\tilde D_{A_0}^A(j=2,\xi(1))
       =\int_0^1du\, u^{j-1}\,D_{A_0}^A(u,\xi(1))\bigg|_{j=2}.
\label{eq:mellin2}
\end{equation}
The other moments can be obtained by differentiation after the exponent $j$
\begin{equation}\label{eq:ulnu}
\la u\ln^i u\ra _{A_0}^A=\frac{d^i}{dj^i}\tilde D_{A_0}^A(j,\xi(1))\bigg|_{j=2}=
\int_0^1du\,u^{j-1}\ln^{i}u\, D_{A_0}^A(u,\xi(1))\bigg|_{j=2}.
\end{equation}
The lower bound in Eqs. (\ref{eq:mellin2}) and
(\ref{eq:ulnu}) has been set to $``0"$ in the limit 
$Q_0\ll E\Theta$, as the threshold region in the integral Eq. (\ref{eq:Nconv})
is exponentially suppressed.
The second ${\cal O}(\alpha_s)$ term
in Eq. (\ref{eq:sv}) follows from the scaling violation in DGLAP
fragmentation functions when one sets $u=1$ in the third argument of $D_{A_0}^A$
in Eq. (\ref{eq:Nconv}); it has been proved in App. D of 
\cite{PAM} not to exceed $5\%$
of the leading contribution, that is why we neglect it hereafter and simply set
$\la u\ra _{A_0}^A=\tilde D_{A_0}^A(j=2,\xi(1))$ as in \cite{Basics,PAM}.

Now Eq. (\ref{eq:decorr}) can be conveniently rewritten in the form
\begin{eqnarray}
\hat N_{A_0}^h(Y_{\Theta_0},Y_{\Theta})&\approx&\sum_{A}\left(\la u\ra _{A_0}^A
+\la u\ln u\ra _{A_0}^A\frac{d\ln N_A^h}{dY_{\Theta}}
+\frac{1}{2}\la u\ln^2 u\ra _{A_0}^A\frac1{N_A^h}
\frac{d^2N_A^h}{d{Y}^2_{\Theta}}\right)
N_A^h(Y_\Theta)\notag\\
&+&{\cal O}(\alpha_s^{3/2})\label{eq:decorrbis}
\end{eqnarray}
and, using Eqs. (\ref{eq:logderN}) and (\ref{eq:seclogderN}), we obtain an
expansion in $\gamma_0\sim \sqrt{\alpha_s}$
\begin{eqnarray}
\hat N_{A_0}^h(Y_{\Theta_0},Y_{\Theta})&
\approx&\sum_{A}\left[\la u\ra _{A_0}^A
+\la u\ln u\ra _{A_0}^A\gamma_0\right.\notag\\
&+&\left.\left(\frac{1}{2}\la u\ln^2 u\ra _{A_0}^A-a_1\la u\ln u\ra _{A_0}^A\right)
\gamma_0^2\right]N_A^h(Y_\Theta)
+{\cal O}(\gamma_0^3)\label{eq:decorrter}
\end{eqnarray}
where the contribution 
$\propto\la u\ra _{A_0}^A$ 
is the leading one (LLA) determined in \cite{Basics}
\begin{eqnarray}
\la u\ra _G^Q&=&\beta(1-e^{-\gamma\xi}), \qquad \la u\ra _G^G=\alpha+\beta e^{-\gamma\xi},\\
\la u\ra _Q^G&=&\alpha(1-e^{-\gamma\xi}), \qquad \la u\ra _Q^Q=\beta+\alpha e^{-\gamma\xi}
\end{eqnarray}
with $\gamma=\frac83C_F+\frac23n_f$, $\alpha=\frac83\frac{C_F}{\gamma}$,
$\beta=\frac23\frac{n_f}{\gamma}$, where $\alpha+\beta=1$ and $\xi=\xi(1)$. 
The functions $\la u\ln u\ra _{A_0}^A$ and $\la u\ln^2u\ra _{A_0}^A$ are
determined from (\ref{eq:ulnu}), as in \cite{PerezMachet}.
Taking $N_c$ and $C_F$ at their QCD values the results still depend on $n_f$.
As this dependence is rather weak, we present the results 
with numerical coefficients only for $n_f=3$
in the appendix.
In the limit $\Theta\to\Theta_0,\ (\xi\to 0)$ the expression  in Eq. (\ref{eq:decorrter})
fulfills the limit Eq. (\ref{subjetlim1}) as the mean values in Eq.
(\ref{eq:ulnu}) vanish and $\la u\ra_{A_0}^A\to \delta_{A_0}^A$. 

\subsection{Multiplicity and color current}
\label{subsec:CC}
The expression  for $\hat N_{A_0}^h$
can be written in compact form as in Eq. (\ref{ColCurDef}) 
introducing the color current.
In the sum over $A$ in Eq. (\ref{eq:decorrter}) when
$A$ is a quark ($A=Q$) we replace
\begin{equation}
N_Q^h=\frac{C_F}{N_c}(1+r_1\gamma_0+\tilde r_2\gamma_0^2)N_G^h,
\quad \tilde r_2=r_1^2+r_2=0.46, \label{eq:nhq}
\end{equation}
according to Eq. (\ref{eq:QGratio}) and after multiplication with the terms
inside the square bracket we keep only terms up to ${\cal O}(\gamma_0^2)$  
so as to respect the NMLLA scheme $(1+\gamma_0+\gamma_0^2)$.
One thus obtains the average multiplicity 
$\hat N_{A_0}^h(Y_{\Theta_0},Y_{\Theta},\lambda)$
of soft hadrons 
within sub-jet angle $ \Theta$ with respect to the
energy flow in terms of the gluon jet multiplicity
 $N_G^h(Y_{\Theta},\lambda)$ 
\begin{equation}\label{eq:Ncc}
\hat N_{A_0}^h(Y_{\Theta_0},Y_{\Theta},\lambda)
\approx\frac{1}{N_c}\la C\ra _{A_0}(Y_{\Theta_0},Y_{\Theta},\lambda)\,
N_G^h(Y_{\Theta},\lambda),
\end{equation}
where $N_G^h(Y_{\Theta},\lambda)$ is given by the solution Eq.   
(\ref{eq:NGhNMLLA}) of the evolution equation (\ref{eq:NGh}) and
$\la C\ra _{A_0}(Y_{\Theta_0},Y_{\Theta},\lambda)$
 ($\equiv \la C\ra _{A_0}(\xi))$ is the average color current of partons 
forming the energy flux
\begin{eqnarray}\label{eq:CA0}
\la C\ra _{A_0}(\xi)\!\!&\!\!=\!\!&\!\!N_c\la u\ra _{A_0}^G(\xi)+C_F\la u\ra _{A_0}^Q(\xi)\\
\!\!&\!\!+\!\!&\!\!\gamma_0(E\Theta)\left[N_c\la u\ln u\ra _{A_0}^G(\xi)+C_F\left(\la u\ln u\ra _{A_0}^Q(\xi)+r_1\la u\ra _{A_0}^Q(\xi)\right)\right]\notag\\
\!\!&\!\!+\!\!&\!\!\gamma_0^2(E\Theta)\left[N_c\left(\frac12\la u\ln^2u\ra _{A_0}^G(\xi)
-a_1\la u\ln u\ra _{A_0}^G(\xi)\right)\right.\notag\\
\!\!&\!\!+\!\!&\!\!\left.C_F\left(\frac12\la u\ln^2 u\ra _{A_0}^Q(\xi)-
(a_1-r_1)\la u\ln u\ra _{A_0}^Q(\xi)+\tilde r_2\la u\ra _{A_0}^Q(\xi)\right)\right].\notag
\end{eqnarray} 

These color currents $\la C\ra _{A_0}$ are also presented in the appendix
with numerical coefficients for $n_f=3$ (see Eqs. (\ref{eq:cg}) and
(\ref{eq:cq})). 
The ratio of the gluon to the quark jet average multiplicity reads
\begin{equation}\label{ref:Ng:Nq}
\frac{\hat N_G^h}{\hat N_Q^h}(\xi)=
\frac{\la C\ra _{G}(\xi)}{\la C\ra _{Q}(\xi)}.
\end{equation}
For this quantity in the limit
$\Theta\to\Theta_0$ ($\xi\to 0$), the appropriate ratio 
$r=\frac{N_G}{N_Q}=r_0(1-r_1\gamma_0-r_2\gamma_0^2$)
is recovered.

\begin{figure}[t]
\begin{center}
\epsfig{file=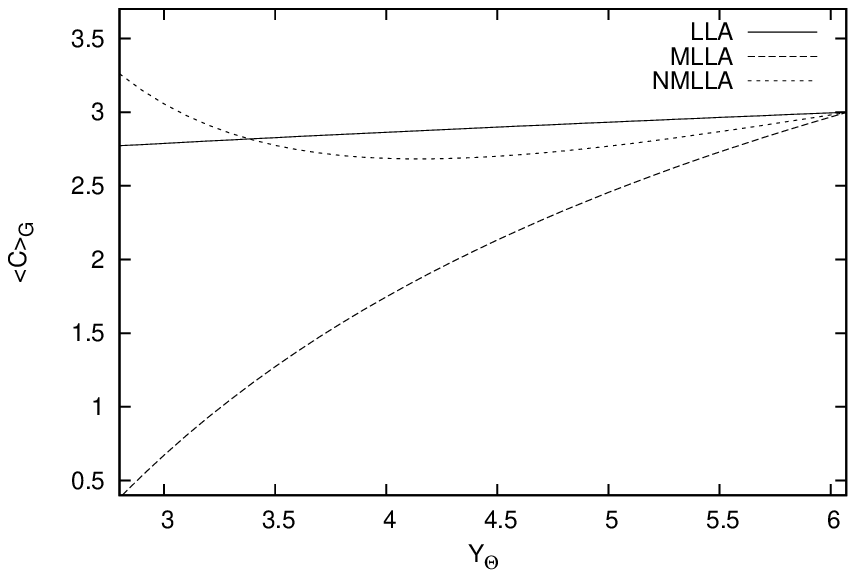,width=7.5truecm}
\qquad
\epsfig{file=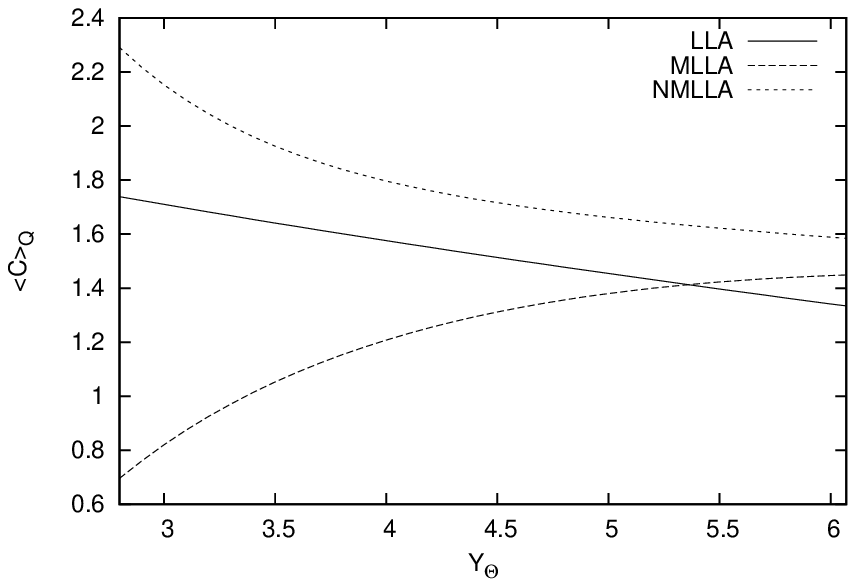,width=7.5truecm}
\caption{Color currents $\la C\ra_G$ (left) and $\la C\ra_Q$ (right)
as a function of $Y_{\Theta}=\ln(E\Theta/Q_0)$
for $Y_\Theta\leq Y_{\Theta_0}=\ln(E\Theta_0/Q_0)$ with $E\Theta_0=100$ GeV
for different approximations (LLA, MLLA, NMLLA) with 
$Q_0=\Lambda_{QCD}=230$ MeV.
\label{fig:cc}
}
\end{center}
\end{figure}

The color currents inside gluon 
and quark jets  $\la C\ra_G$ and  $\la C\ra_Q$ are displayed
in Fig.\,\ref{fig:cc} for a particular jet virtuality $E\Theta_0=100$ GeV
in the LLA, MLLA and NMLLA. 
For simplicity we also use $Q_0=\Lambda_{\rm QCD}$ ($\lambda=0$) 
which is known to be a good approximation for hadron spectra
(``limiting spectrum'').
As expected from Eqs. (\ref{eq:CA0}) and (\ref{eq:cg}), 
for $\Theta\to\Theta_0\ (\xi\to 0)$, 
all curves for $\la C\ra_G(Y_{\Theta})$ 
coincide at $\la C\ra_G(Y_{\Theta_0})=N_c$, where $N_c$ is the color
factor corresponding to the splitting of one gluon 
(in this case the one initiating the jet) into two
other gluons.  
In the same limit $\Theta\to\Theta_0$, the color current
$\la C\ra_Q$ in LLA approaches $C_F$, where $C_F$ is the color factor 
corresponding to the splitting of a quark into a quark
and a gluon. 
The mismatch at 
$\Theta=\Theta_0$ between the MLLA and the LLA curves comes from the
$\gamma_0$-expansion of $r=N_G/N_Q$ and equals $C_Fr_1\gamma_0$, 
while the difference between the LLA curve and the
NMLLA curve is increased by the ${\cal O}(\gamma_0^2)$ correction
and reaches $C_F(r_1\gamma_0+\tilde r_2\gamma_0^2$). 

The role of ${\cal O}(\gamma_0)$ and ${\cal O}(\gamma_0^2)$ corrections to
the LLA color current $\la C\ra_{A_0}$ in Eq. (\ref{eq:CA0}) 
is quite important 
and the expansion in $\gamma_0$ is seen in Fig. \ref{fig:cc} 
to oscillate if higher order terms
are incorporated. Indeed, the terms
$\gamma_0^i\la u\ln^i u\ra\propto(-\gamma_0)^{i}$ change 
sign as one goes from the LLA ($i=0$) to MLLA ($i=1$) and to the NMLLA ($i=2$).
Because of the running coupling the $\gamma_0$-expansion 
($\gamma_0\sim\sqrt{\alpha_s}$) converges at high energies but may diverge
at low energies. At present energies ($\gamma_0\sim 0.5$ at LEP)
the $\gamma_0$ expansion is converging rather slowly and one may ask for yet
higher order terms. Such calculations would also require higher order terms
in the expansion of the ratio $r$ and its derivatives which are not
available. As an exercise, we calculated 6 terms beyond LLA for the gluon
current numerically while keeping the expansion of $r$ at second order
in  $\gamma_0$.
Then the solutions keep oscillating with a new solution falling in between
the two previous ones. The trend is towards a solution closer to NMLLA than
to MLLA. 

Starting with the isolated jet at $\Theta_0\ (Y_{\Theta_0})$ 
the sub-jets at smaller angles $\Theta\ (Y_{\Theta})$ 
may evolve from intermediate partons of different color. Therefore,
one expects that sub-jets in a gluon jet have lower multiplicity
than isolated gluon jets at the same angle ($Y_{\Theta}$) 
and the opposite for a quark
jet. This behavior is in fact born out by the 
 MLLA and NMLLA curves for
$\la C\ra_G$ which show the same trends as the LLA term: it is 
increasing with $Y_\Theta$ at the higher energies.
On the other hand, the quark color current $\la C\ra_Q$ is decreasing in
LLA and in the NMLLA while for the intermediate approximation MLLA
this behavior is not yet reached. 
Asymptotically ($\gamma_0\to0$) all curves have to approach either 
$\la C\ra_G=N_c$ or $\la C\ra_Q=C_F$. 

\subsection{Results on multiplicities at small angles}

\begin{figure}[t]
\begin{center}
\epsfig{file=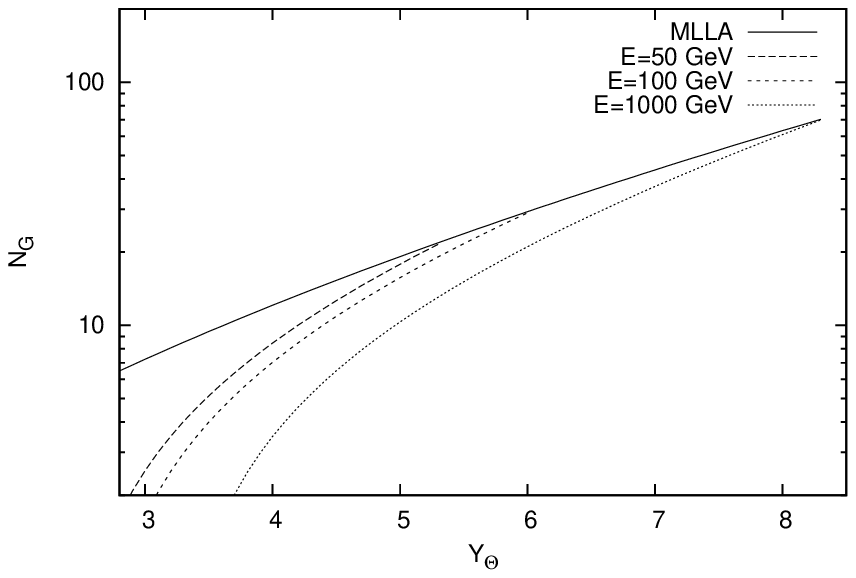,width=7.5truecm}
\qquad
\epsfig{file=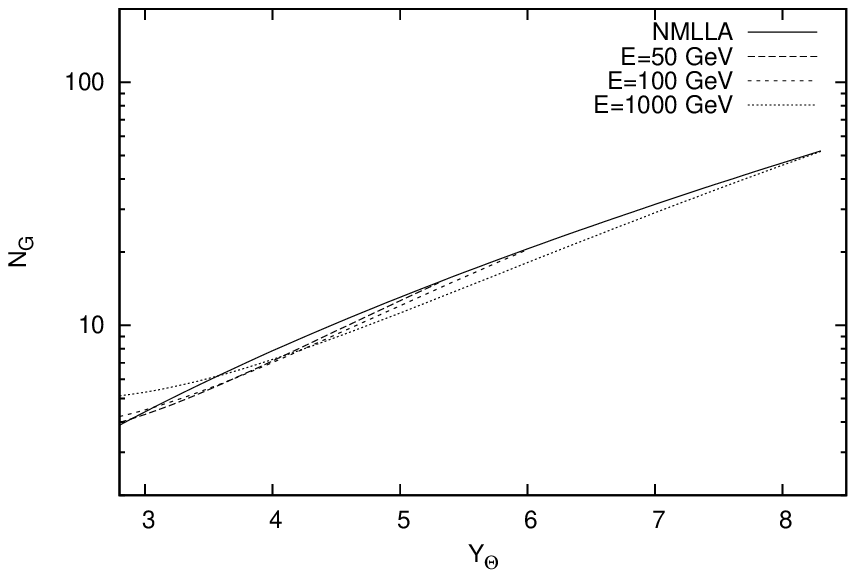,width=7.5truecm}
\caption{\label{fig:MultG}Multiplicity in gluon jets 
as function of the variable
$Y=\ln(E\Theta/Q_0)$: the full line represents multiplicity of isolated jet
$N_G(E\Theta_0)$ at fixed opening angle 
$\Theta=\Theta_0=1$ varying the energy $E$, the other curves represent
sub-jet multiplicities $\hat N_G(E,\Theta_0,\Theta)$ at
fixed energies $E=50,100,1000$ GeV varying the angles $\Theta$; 
left panel:in MLLA, right panel: in NMLLA ($Q_0=\Lambda_{\rm QCD}=0.23$ GeV).
}
\end{center}
\end{figure}

\begin{figure}[t]
\begin{center}
\epsfig{file=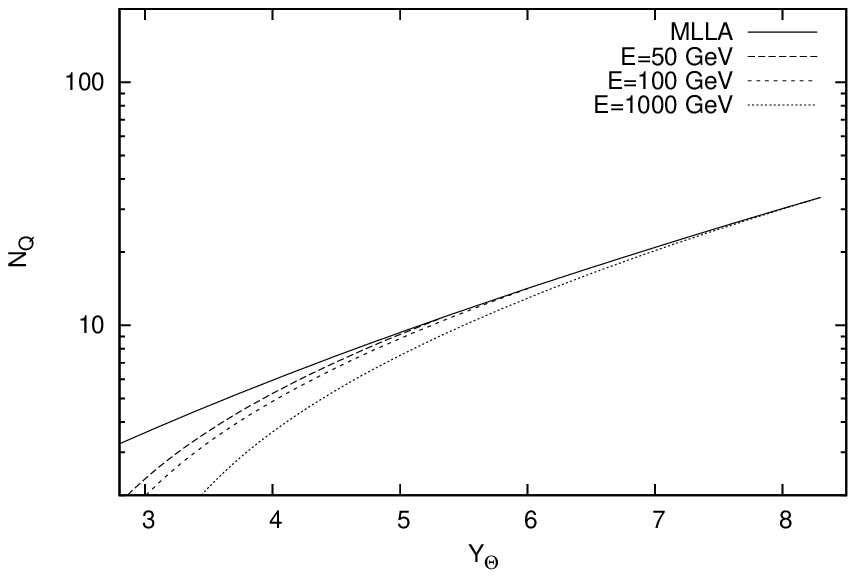,width=7.5truecm}
\qquad
\epsfig{file=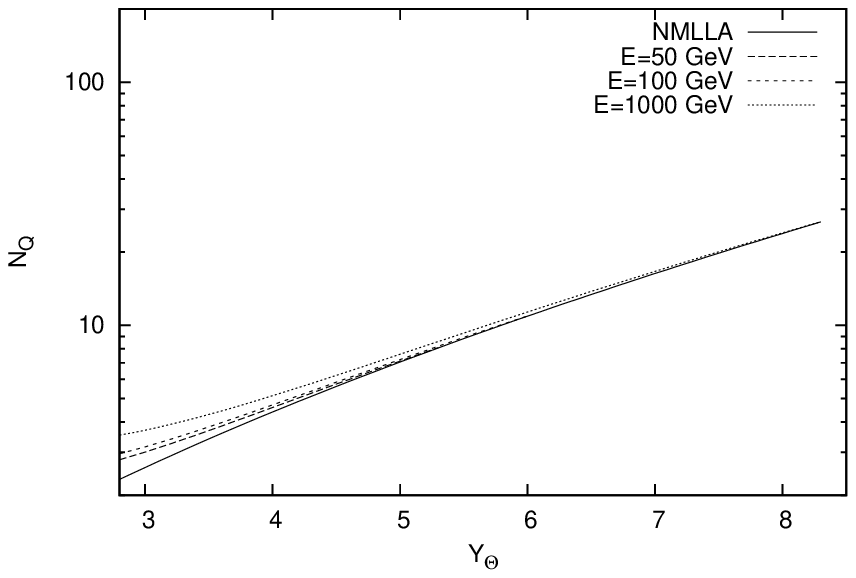,width=7.5truecm}
\caption{\label{fig:MultQ}Multiplicities  as in Fig. \protect\ref{fig:MultG},
but for quark jets.}
\end{center}
\end{figure}

Next we study the consequences for the behavior of multiplicities
at full angles $\Theta_0\sim 1$ and for sub-jets at reduced angles $\Theta$.
Results for multiplicities using our formulae for color currents Eqs.
(\ref{eq:Ncc}) and (\ref{eq:CA0}) are
shown in Figs. \ref{fig:MultG} and \ref{fig:MultQ}
for MLLA (left panels) and NMLLA (right panels).
The full lines show the particle
multiplicity $N_{A_0}$ for an isolated $A_0$ jet as function of $Y\equiv
Y_{\Theta_0}=\ln(E\Theta_0/Q_0)$, 
i.e. as function of the jet 
energy $E$ for the gluon jet in Fig.  \ref{fig:MultG} and for the quark jet 
in Fig.  \ref{fig:MultQ}, according to Eqs. (\ref{eq:rgq}) and
(\ref{eq:NGhNMLLA}). The hadronization constant 
has been taken as $K'=0.2$ according to
\cite{DreminGary}.
\begin{figure}[h]
\begin{center}
\epsfig{file=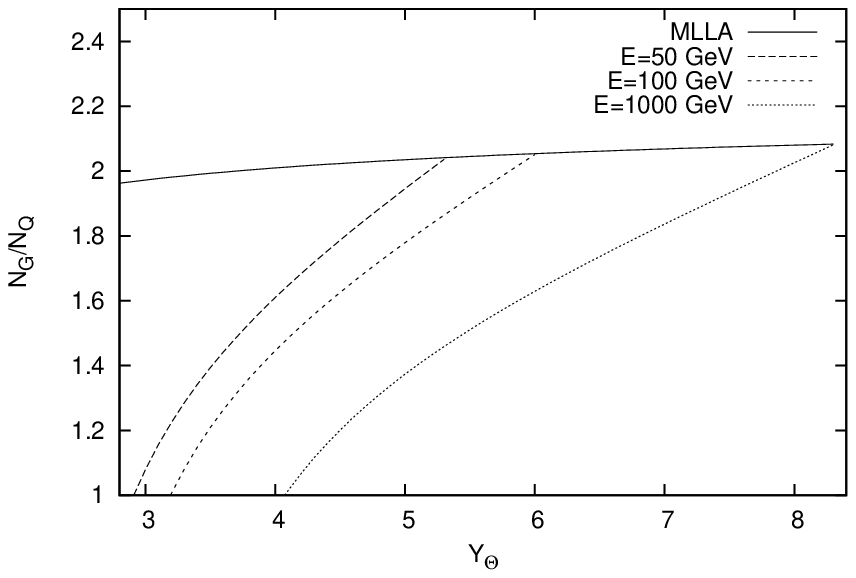,width=7.5truecm}
\qquad\epsfig{file=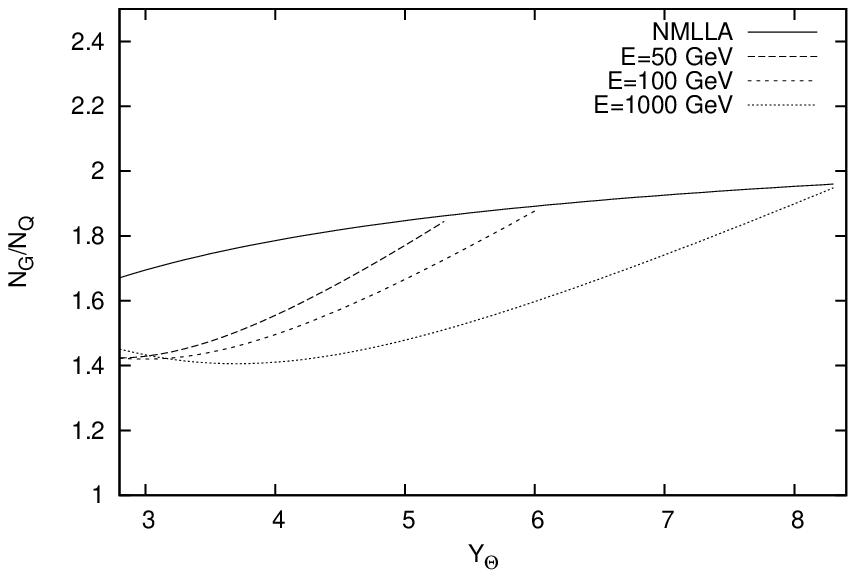,width=7.5truecm}
\caption{\label{fig:MultRat}Multiplicity ratio $N_G/N_Q$ for jets and sub-jets
as function of energy variable
$Y=\ln(E\Theta/Q_0)$ as in Fig. \protect\ref{fig:MultG}.
}
\end{center}
\end{figure}
Also shown in these figures are the multiplicities $\hat N^h_{A_0}$ 
for the reduced opening
angle $\Theta$ as function of $Y_\Theta=\ln(E\Theta/Q_0)$ for
different fixed energies $E$ but variable sub-jet angles
$\Theta$. For $\Theta \to \Theta_0$
these curves approach the full multiplicities at $Y=Y_{\Theta_0}$. 
In the simple model with
the scaling of multiplicities with $E\Theta$ in Eq. (\ref{eq:scale}) 
all these curves
would coincide; the existence of intermediate processes and non-trivial
color currents yields the predicted scale breaking. The mixing of quark and
gluon jets will reduce the multiplicity in the gluon jet and increase it in
the quark jet at sufficiently high energies. 
This property is reproduced in our calculations except for $\hat N_{A_0}^h$
in MLLA where the appropriate high energy regime is not yet reached.

In case of the ratio $r=N_G/N_Q$ the effects from gluon and quark jets go in
the same direction and yield the observable scale breaking 
effects of Fig. \ref{fig:MultRat} for the small cone measurements.
For the LEP energy range ($Y\sim 5$)
the ratio for large jet opening angle $\Theta_0$ is $r\sim 2$ in MLLA and
$r\sim 1.8$ in NMLLA. This is
still larger than the experimentally observed $r\sim 1.5$
\cite{OPALrgq,DELPHIrgq} which indicates the slow convergence of the
$\gamma_0\sim \sqrt{\alpha_s}$ expansion of multiplicity. As emphasized in Sect. 2
the numerical treatment of the evolution equation removes this discrepancies
largely.
The effect from intermediate processes in the new definition of multiplicity
amounts to about 20\% in NMLLA at a reduced energy scale $Y\sim Y_0-2$. 

\section{Conclusions}

We have studied a definition of jet axis which is based on a 2-particle
correlation where the jet axis corresponds to the energy weighted direction
of particles in a given cone $\Theta_0$. In this way smaller angles (or transverse
momenta) can be meaningfully determined.
We studied especially the 
effects related to a measurement of multiplicities where
we have added MLLA and NMLLA corrections to the known LLA results
and studied phenomenological consequences numerically.
  
The main new effect from 2-particle correlations is the appearance of the
``color current'' in the result which reflects the possibility of
intermediate quark-gluon processes. This effect can be seen by the scale
breaking between jets and sub-jets at the same scale $E\Theta$ but different
energies $E$ and opening 
angles $\Theta$. Typical effects are of the order of 20\%.

The expansion in $\sqrt{\alpha_s}$ which is used in the calculation of color
currents is rather slowly converging and MLLA in the considered energy range
is not satisfactory whereas NMLLA shows the expected behavior at the
qualitative level. For a more quantitative treatment higher order terms
have to be calculated but this requires also a treatment of the multiplicity
ratio $r$ at higher orders which is not yet available. Alternatively,
numerical methods and Monte Carlo calculations may be accessible.

As an interesting application, one could learn from a single type of jet 
about properties of the other type by the study of sub-jets of lower
energies (for
example, from a high energy gluon jet at LHC about quark jets of lower
energies). Other applications concern
behavior of spectra at small transverse momenta or low energy gluon jets.


\newpage

\begin{Large}
{{\bf Appendix:
Expressions for $\boldsymbol{\la u\ln^i u\ra _{A_0}^A}$,
and the color currents $\boldsymbol{\la C\ra _{A_0}}$
}}
\end{Large}

\label{sec:colorcurr}

Setting $\xi\equiv\xi(1)$ we obtain the MLLA correction from 
Eq. (\ref{eq:ulnu}) for $i=1$ and $n_f=3$

\begin{equation}
\la u\ln u\ra _G^G(\xi)=-0.875023-7.80247 \xi+e^{-5.55556 \xi } (0.875023-1.65883 \xi ),
\end{equation}

\begin{equation}
\la u\ln u\ra _G^Q(\xi)=0.326411-4.38889 \xi +e^{-5.55556 \xi } (-0.326411+1.65883 \xi ),
\end{equation}

\begin{equation}
\la u\ln u\ra _Q^G(\xi)=0.100287-7.80247 \xi +e^{-5.55556 \xi } (-0.100287+2.94903 \xi ),
\end{equation}

\begin{equation}
\la u\ln u\ra _Q^Q(\xi)=0.875023-4.38889 \xi+e^{-5.55556 \xi } (-0.875023-2.94903 \xi );
\end{equation}

for the NMLLA correction with $i=2$ we also write the result for $n_f=3$ 

\begin{eqnarray}
\la u\ln^2u\ra _G^G(\xi)&=&7.64366 e^{-5.55556 \xi } (-1.0829+\xi ) (-0.146177+\xi )\notag\\
&+&95.1228 (-0.0286673+\xi ) (0.443708+\xi ),\\
\notag\\
\la u\ln^2u\ra _G^Q(\xi)&=&53.5066 (-0.117852+\xi ) (0.159854+\xi )\notag\\
&-&7.64366 e^{-5.55556 \xi } (-0.744713+\xi ) (0.177083+\xi ),\\
\notag\\
\la u\ln^2u\ra _Q^G(\xi)&=&95.1228 (-0.0746644+\xi ) (0.239705+\xi )\notag\\
&-&13.5887 e^{-5.55556 \xi } (-0.495132+\xi ) (0.253032+\xi ),\\
\notag\\
\la u\ln^2u\ra _Q^Q(\xi)&=&1.20995+13.5887 e^{-5.55556 \xi } (-0.155024+\xi ) (0.574368+\xi )\notag\\
&+&\xi  (-11.1292+53.5066 \xi ).
\end{eqnarray}

Substituting the former results in (\ref{eq:CA0}) and setting
$$
\gamma_0(E\Theta)=\gamma_0(E\Theta_0)e^{2N_c\beta_0\xi},\quad
\gamma_0^2(E\Theta)=\gamma_0^2(E\Theta_0)e^{4N_c\beta_0\xi}
$$
yields the color currents, evaluated again for $n_f=3$,
\begin{eqnarray}\label{eq:cg}
\la C\ra _G(\xi)&=&2.4+0.6\,e^{-5.55556\,\xi }\\
&+&\gamma_0(E\Theta_0)\,e^{-1.05556\,\xi } \left(2.10105+e^{5.55556 \xi }
(-2.10105-29.2593 \xi )-2.76472 \xi \right)\notag\\
&+&\gamma_0^2(E\Theta_0)\left(
6.36971\,e^{3.44444\,\xi } (-1.39557+\xi ) (-0.176893+\xi )\notag\right.\\
&+&\left.178.355\,e^{9\,\xi } (-0.0219197+\xi ) (0.402217+\xi )\right)
+{\cal O}(\gamma_0^3),\notag
\end{eqnarray}
\begin{eqnarray}\label{eq:cq}
\la C\ra _Q(\xi)&=&2.4-1.06667\,e^{-5.55556\,\xi }\\
&+&\gamma_0(E\Theta_0)\,
\left(e^{-1.05556\,\xi }\left(-1.30969+e^{5.55556\,\xi } (1.55636-29.2593 \xi )+4.91505 \xi \right)
\right)\notag\\
&+&\gamma_0^2(E\Theta_0)\left(178.355\,e^{9\,\xi }
(-0.0527303+\xi )(0.183028+\xi )-11.3239\,e^{3.44444 \xi }\notag\right.\\
&&\left. (-0.832991+\xi ) (0.247506+\xi )\right)+{\cal O}(\gamma_0^3).\notag
\end{eqnarray}

After replacing $\xi$ by $\xi(u=1)$ in (\ref{eq:xiu}) one obtains   
the functions
$\la C\ra _{G,Q}(Y_{\Theta_0},Y_\Theta,\lambda)$ 
involved in (\ref{eq:Ncc}).


\null




\newpage

\begin{thebibliography}{50}
%
\bibitem{Basics}
Yu.L. Dokshitzer, V.A. Khoze, A.H. Mueller \& S.I. Troyan, Basics of
Perturbative QCD, Editions Fronti\`eres, Paris (1991).

\bibitem{LPHD}
Ya.I. Azimov, Yu.L. Dokshitzer, V.A. Khoze \& S.I. Troian,
Z. Phys. {\bf C} 27 (1985) 65.

Yu.L. Dokshitzer, V.A. Khoze \& S.I. Troian,
J. Phys. {\bf G} 17 (1991) 1585.

\bibitem{KhozeOchs}
V.A. Khoze \& W. Ochs;
Int. J. Mod. Phys. {\bf A} 12 (1997) 2949.

\bibitem{MultTheory}
A.H. Mueller, Nucl. Phys. {\bf B} 241 (1984) 141; Erratum ibid., {\bf B} 241 (1984) 141.

\bibitem{MALAZA}
E.D. Malaza \& B.R. Webber, Phys. Lett. {\bf B} 149 (1984) 501.

\bibitem{DREMIN}
I.M. Dremin, V.A. Nechitailo, Mod. Phys. Lett. {\bf A} 9 (1994) 1471;
JETP Lett. {\bf 58} (1993) 945.

\bibitem{OPALrgq}
G. Abbiendi et al., [OPAL Collaboration], Phys. Rev. {\bf D} 69 (2004) 032002. 

\bibitem{DELPHIrgq}
J. Abdallah et al. [DELPHI Collaboration], Eur. Phys. J. {\bf C} 44 (2005) 311.

\bibitem{DreminGary}
I.M. Dremin \& J.W. Gary, Phys. Rep. {\bf 349}
(2001) 301.

\bibitem{TevMult}
D. Acosta et al., Phys. Rev. Lett. {\bf 94} 171802 (2005). 

\bibitem{TevRap}
A.N. Safonov (for CDF Collaboration), Nucl. Phys. {\bf B} (Proc. suppl.) 86
(2000) 55.

\bibitem{DDT}
Yu.L. Dokshitzer, D.I. Dyakonov \& S.I. Troyan, Phys. Rep. {\bf 58} (1980) 270.

\bibitem{PerezMachet}
R. Perez-Ramos \& B. Machet, JHEP {\bf 04} (2006) 043.



\bibitem{PAM} R. Perez-Ramos, F. Arl\'eo \& B. Machet, arXiv:0712.2212 [hep-ph],
Phys. Rev. {\bf D} in press;
F. Arl\'eo, R. Perez-Ramos \& B. Machet, Phys. Rev. Lett. {\bf 100} (2008) 052002  . 

\bibitem{CDF}
S. Jindariani,  A. Korytov \& A. Pronko: ``$k_\perp$ Distributions of
Particles in Jets at CDF'',
CDF report CDF/ANAL/JET/PUBLIC/8406 (March 2007),\hfill\break
www-cdf.fnal.gov/physics/new/qcd/ktdistributions\_06/cdf8406\_Kt\_jets\_public.ps


\bibitem{LupiaOchs}
S. Lupia \& W. Ochs, Phys. Lett. {\bf B} 418 (1998) 214.

\bibitem{HERWIG}
G. Corcella et al., JHEP {\bf 01} (2001) 010.


\end{thebibliography}
\end{document}